\documentstyle[12pt]{article}

\normalbaselines
\begin{document}
\begin{center}
\vspace*{1.0cm}

{\LARGE{\bf  Symmetry and Covariance of  Non-relativistic Quantum Mechanics}}

\vskip 1.5cm

{\large {\bf M. Omote$^{1)}$ and S. Kamefuchi$^{2)}$ }}

\vskip 0.5 cm

Department of Physics, Keio University, Hiyoshi, Yokohama$^{1)}$ 

Atomic Energy Research Institute, Nihon University, Kanda-Surugadai, Tokyo$^{2)}$

\end{center}

\vspace{1 cm}

\begin{abstract}
On the basis of a 5-dimensional form of space-time transformations non-relativistic quantum mechanics is reformulated in a manifestly covariant manner. The resulting covariance resembles that of the conventional relativistic quantum mechanics.

\end{abstract}

\vspace{1 cm}

\section{ Introduction}
In contrast with relativistic quantum mechanics (RQM) non-relativistic quantum mechanics (NRQM), in the conventional formulation, lacks manifest covariance with respect to general, space-time coordinate transformations. The main reason for this is that basic quantities such as fields or Schr$\ddot{{\rm o}}$dinger probability amplitudes are taken there not to be vector, but to be projective representations of the 4-dimensional Galilei group $G_{4}$ or of the group of more general transformations. The reason can further be traced to the fact that Lagrangians of relevant systems recover invariance under those transformations, only when some time-derivative terms are excepted.

In our previous paper [1],  we have shown, however, that NRQM in inertial systems can be formulated in a manifestly covariant manner, provided  a 5-dimensional form $G_{5}$ is adopted for Galilei transformations [2].  In the present report we generalize $G_{5}$ to those transformations $G_{5}^{\prime}$ which connect inertial with non-inertial systems, and thereby finds a manifestly covariant formulation of NRQM, being valid for general coordinate systems, inertial or non-inertial. The formulation, in fact, runs quite parallel with that of RQM.

Interaction terms can be introduced in a way similar to RQM. In particular, those with external electromagnetic or gravitational fields are generated by certain replacements. The resulting equations for non-inertial cases contain terms corresponding  to inertial forces, and this enables us to discuss, e.g., the equivalence principle solely within the framework of NRQM.

\section{$G_{5}$ and Manifest Covariance }

The basic ideas of [1] are as follows. The Lagrangian ${\cal L}= (1/2)m \dot{\vec{x}}^{2}$ of a free particle of mass m is not   invariant under $G_{4}$:  $\vec{x} \to \vec{x}^{\prime}=R\vec{x}-\vec{v}t, t^{\prime}=t$ with $R^{tr}R={\rm I}$. However, $\bar{{\cal L}}\equiv {\cal L}-m \dot{s}$ remains invariant, provided the new variable $s$ transforms under $G_{4}$ as $s \to  s^{\prime}=s+f$ with $f(\vec{x},t) \equiv (R\vec{x})\cdot \vec{x}+(1/2)\vec{v}^{2}t.$ Thus, instead of $G_{4}$ we employ $G_{5}$, a central extension of $G_{4}$, such that 
\begin{equation}
\begin{array}{l}
x^{\prime i}=R^{i}~_{j}~x^{j}-\frac{v^{i}}{u}x^{4},\hspace{1cm} x^{\prime 4}=x^{4},  \\
x^{\prime 5}=x^{5}-\frac{v_{i}}{u}(R^{i}~_{j} x^{j})+\frac{1}{2}\frac{\vec{v}^{2}}{u^{2}}x^{4} ,
\end{array}
\end{equation}
where $(x^{1},x^{2},x^{3}) \equiv \vec{x},~x^{4} \equiv ut, x^{5} \equiv s/u$ with $[u]=[v]$, or $x^{\prime \mu}\equiv \Lambda^{\mu}~_{\nu} x^{\nu}$. Under (1) $\eta_{\mu \nu}x^{\mu}x^{\nu}=\vec{x}^{2}-2ts$ is invariant, with $\parallel \eta_{\mu \nu}\parallel=\parallel \eta^{\mu \nu}\parallel$ such that $\eta_{ij}=\delta_{ij} (i,j=1,2,3), g_{45}=g_{54}=-1$ and others $=0$. Note that $\eta_{\mu\nu}x^{\mu}x^{\nu}$ with $x_{\pm}\equiv (x^{4} \pm x^{5})/\sqrt{2}$ leads to $\vec{x}^{2}+\vec{x}_{-}^{2}-x_{+}^{2}$, and $\eta_{\mu\nu}p^{\mu}p^{\nu}=0$ with $p^{\mu} \equiv (\vec{p},mu,E/u)$ to $E=\vec{p}^{2}/2m$.

Our basic assumption then is that NRQM for free systems be invariant under $G_{5}$. Thus for a scalar field $\phi(x)$ a  Klein-Gordon type equation
\begin{equation}
\eta^{\mu\nu}\partial_{\mu}\partial_{\nu} \phi(x)=0
\end{equation}
with a subsidiary condition  
\begin{flushright}
$(i\hbar \partial_{5}-mu)\phi(x)=0  \hspace{5cm}(2^{\prime})$
\end{flushright}
results, for $\phi(x)=\exp (-ims/\hbar)~ \psi(\vec{x},x^{4})$, in 
\begin{equation}
i\hbar\frac{\partial}{\partial t}\psi(\vec{x},t) = -\frac{\hbar^{2}}{2m}\vec{\nabla}^{2}\psi(\vec{x},t)~. 
\end{equation}
Notice that $\phi^{\prime}(x^{\prime})=\phi(x)$ leads immediately to $\psi^{\prime}(\vec{x}^{\prime},t^{\prime})=\exp (imf/\hbar)\psi(\vec{x},t)$ with $\psi^{\prime}(\vec{x}^{\prime},t^{\prime})$ similarly defined. Likewise, for a spinor field $\chi(x)$ we assume a Dirac-type equation
\begin{equation}
\gamma^{\mu}\partial_{\mu}\chi (x) =0
\end{equation}
together with a subsidiary condition of the same form as (2$^{\prime}$). Here, the $\gamma$-matrices satisfying  $\gamma^{\mu}\gamma^{\nu} + \gamma^{\nu}\gamma^{\mu} = 2\eta^{\mu \nu} $ have the 4$\times$ 4 irreducible representation. Under $G_{5}$ $\chi(x)$ transforms as $\chi^{\prime}(x^{\prime})=T\chi(x)$, where $T^{-1}\gamma^{\mu}T=\Lambda^{\mu}~_{\nu}\gamma^{\nu}$. Substituting 
\begin{equation}
\chi(x)=  \exp\Bigl(-\frac{ims}{\hbar}\Bigr)\left(
\begin{array}{c}
\psi_{1}(\vec{x},x^{4})  \\
\psi_{2}(\vec{x},x^{4})
\end{array}
\right) 
\end{equation}
into (4), we find that only the 2-component spinor $\psi_{1}$ is independent, and satisfies the equation of the same form as (3). The extension of the above arguments to the case of Bargmann-Wigner fields of higher spins is straightforward.

By using vector representations of $G_{5}$ such as exemplified in the above we can construct a manifestly covariant field theory. It should be noted here that when a field $\psi(\vec{x},t)$ satisfies a linear equation of the Schr$\ddot{{\rm o}}$dinger type such as (3), the probability amplitude $\varphi(\vec{x},t)$ for a particle, resulting from the quantized $\psi$, does also satisfy the same equation. Thus, the same covariance arguments apply as well to $\varphi$.

\section{General Coordinate Transformations}

Generalizing $G_{4}$ we consider space-time transformations $G_{4}^{\prime}$ such as $\vec{x}^{\prime}=R(t)\vec{x}+\vec{A}(t),~t^{\prime}=t$, where $R$ and a vector $\vec{A}$ are taken to be $t$-dependent. The corresponding transformation rule of $s$ can again be found by observing how ${\cal L}$ transforms under $G^{\prime}_{4}$. Thus, instead of (1) we now consider $G_{5}^{\prime}$ such that 
\begin{equation}
\begin{array}{l}
x^{\prime i}=R^{i}~_{j}x^{j}+A^{i},   \hspace{1cm} x^{\prime 4}=x^{4},  \\
x^{\prime 5}=x^{5}+\check{\tilde{A}}_{j}x^{j}+\frac{1}{u}\tilde{A}_{j}\dot{\tilde{A}}~^{j}-\frac{1}{2u^{2}}\int_{0}^{x^{4}}\dot{\tilde{A}}_{j}(\tau )\dot{\tilde{A}}~^{j}(\tau)d \tau,   
\end{array} 
\end{equation}
where $\tilde{A}_{i} \equiv R^{j}~_{i}A_{j}$. Obviously, the transformation converts, in general, an inertial system S$_{0}$ (coordinates : $x^{\mu}$) to a non-inertial system S (coordinates : $x^{\prime \mu}$).

In S the metric tensor $g^{\prime \mu\nu}$ is given by
\begin{equation}
\begin{array}{lll}
g^{\prime ij}=\delta^{ij}, & g^{\prime i4}=0, & g^{\prime i5}=-\frac{1}{u}\dot{R}^{i}~_{j}R_{k}~^{j}x^{\prime k},  \\
g^{\prime 44}=0, & g^{\prime 45}=-1, & g^{\prime 55}=-\frac{2}{u^{2}}R_{ji}~\ddot{\tilde{A}}~^{i}~x^{\prime j};
\end{array}
\end{equation}
and the affine connection $\Gamma^{\prime \lambda}_{\mu \nu}$ by
\begin{equation}
\begin{array}{ll}
\Gamma^{\prime i}_{4j}=\frac{1}{u}R^{i}~_{k}\dot{ R}_{j}~^{k},& \Gamma^{\prime i}_{44}=\frac{1}{u^{2}}R^{i}~_{j}\ddot{ R}_{k}~^{j}x^{\prime k}-\frac{1}{u^{2}}R^{i}~_{j}\ddot{\tilde{A}} ~^{j},  \\
\Gamma^{\prime 5}_{4i}=-\frac{1}{u^{2}}R_{ij}\ddot{\tilde{A}}~^{j}, & \Gamma^{\prime 5}_{44 }=-\frac{2}{u^{3}}\dot{R}_{ij}~\ddot{\tilde{A}}~^{j}~x^{\prime i}-\frac{1}{u^{3}}~R_{ij}\stackrel{\cdots}{\tilde{A}}~^{j}~x^{\prime i},   
\end{array}
\end{equation}
\[{\rm others}=0 . \]

\section{Generally Covariant Field Equations}

In order to generalize the field equation in S$_{0}$ to those in S we have only to follow the procedure employed in going from special to general relativity.

Thus, for the scalar field $\phi^{\prime}(x^{\prime})$ in S the equation (2) is changed to $g^{\prime \mu \nu}{\cal D}^{\prime}_{\mu}{\cal D}^{\prime}_{\nu} \phi^{\prime}(x^{\prime})=0$, where ${\cal D}^{\prime}_{\mu}$ is the covariant derivative. Since, however, $g^{\prime \mu \nu}\Gamma^{\prime \lambda}_{\mu \nu}=0$ from (7) and (8), the above equation reduces simply to
\begin{equation}
g^{\prime \mu \nu}\partial^{\prime}_{\mu}\partial^{\prime}_{\nu} \phi^{\prime}(x^{\prime})=0.
\end{equation}
On the other hand, the subsidiary condition takes the same form as (2$^{\prime}$) because of  $\partial^{\prime}_{5}=\partial_{5}$. Defining $\psi^{\prime}(\vec{x}^{\prime},t^{\prime})$ from $\phi^{\prime}(x^{\prime})$ in the same way as before we find from (9) and (7) that 
\begin{equation}
i\hbar\frac{\partial}{\partial t^{\prime}}\psi^{\prime}(\vec{x}^{\prime},t^{\prime})=\Bigl(-\frac{\hbar^{2}}{2m}\vec{\nabla}^{\prime 2}+{\cal H}^{\prime}_{inert}\Bigr)\psi^{\prime}(\vec{x}^{\prime},t^{\prime}) ,
\end{equation}
\begin{equation}
{\cal H}^{\prime}_{inert} \equiv -mR_{jk}\ddot{\tilde{A}}~^{k}x^{\prime j}-i\hbar\dot{R}^{\ell}~_{j}R_{k}~^{j}x^{\prime k}\partial_{\ell}^{\prime} .
\end{equation}

For the spinor field $\chi^{\prime}(x^{\prime})$, to be regarded as a scalar under (6), the equation (4) is generalized to 
\begin{equation}
 \gamma^{\prime \mu}(x^{\prime})\bigl(\partial_{\mu}^{\prime}+\Gamma_{\mu}^{\prime}(x^{\prime})\bigr) \chi^{\prime}(x^{\prime})=0 ,
\end{equation}
whereas the form of the subsidiary condition remains unchanged. The $\gamma^{\prime}$-matrices and the spin connection $\Gamma^{\prime}_{\mu}(x^{\prime})$ are given, in terms of the f$\ddot{{\rm u}}$nfbein $h^{\mu}_{a}(x^{\prime})$, as
\begin{eqnarray}
\gamma^{\prime \mu}(x^{\prime}) &=&h^{\mu}_{a}(x^{\prime})~\gamma^{a}, \nonumber    \\
\Gamma_{\lambda}^{\prime}(x^{\prime})&=& \frac{1}{8} ~[\gamma^{a},\gamma^{b}]~g^{\prime}_{\mu\nu}(x^{\prime}) h^{\mu}_{a}(x^{\prime}) {\cal D}_{\lambda}^{\prime}h^{\nu}_{b}(x^{\prime}) .
\end{eqnarray}
As usual, we have also  $g^{\prime \mu \nu}(x^{\prime})=h^{\mu}_{a}(x^{\prime})h^{\nu}_{b}(x^{\prime})\eta^{ab}$ and $\gamma^{\prime \mu}\gamma^{\prime \nu}+\gamma^{\prime \nu}\gamma^{\prime \mu}=2g^{\prime \mu \nu}(x^{\prime})$.

By construction (12) is invariant not only under (1), but also under local Galilei transformations $\chi^{\prime}(x^{\prime}) \to \tilde{T}(x^{\prime})\chi^{\prime}(x^{\prime})$ , where 
 $\tilde{T}(x^{\prime})$ corresponds to $x^{\prime \mu}$-dependent $\Lambda^{a}~_{b}$  with $\mu \ne 5$ . Incidentally, the latter is a kind of non-Abelian gauge transformations, and $\Gamma^{\prime}_{\mu}$ a gauge dependent quantity.

When explicitly rewritten in S, (12) takes the following forms. We consider two choices of $h^{\mu}_{a}$ (or of the gauge). 1) We take $h^{\mu}_{a}(x^{\prime})=\partial x^{\prime \mu}/\partial x^{a}$. In this case ${\cal D}^{\prime}_{\lambda} h^{\nu}_{b}(x^{\prime})=0$, hence $\Gamma^{\prime}_{\mu}(x^{\prime})=0$. Then, for the function $\psi^{\prime}_{1}(\vec{x}^{\prime},t^{\prime})$, defined in a way similar to (5), there holds the equation, having the same form as (10) with (11). 2) Taking a $t^{\prime}$-dependent orthogonal matrix $\tilde{R}$, we adopt $h^{\prime i}_{j}=\tilde{R}_{j}~^{k}h^{i}_{k}, h^{\prime 5}_{i}= \tilde{R}_{i}~^{j}h^{5}_{j}$, and $h^{\prime \mu}_{a}=h^{\mu}_{a}$ for others. In this case, on the right hand side of the equation for $\psi^{\prime}_{1}(\vec{x}^{\prime},t^{\prime})$ an additional term ${\cal H}^{\prime}_{{\rm spin}}$ appears , such as
\begin{equation}
{\cal H}^{\prime}_{{\rm spin}}\equiv \frac{\hbar}{4}\dot{\tilde{R}}~^{\ell}~_{j} \tilde{R}^{kj} \epsilon_{k\ell m} \sigma_{m}
\end{equation}
with $\sigma_{m}$'s being Pauli matrices.

For the case of Bargmann-Wigner fields the results are basically the same as above. Needless to say, (11) and (14) provide the general expressions for inertial force potentials.

\section{Further Remarks}

{\bf a}) Interactions can be introduced to (2) or (4) by adding $G_{5}$-symmetry conserving or violating terms. In particular,  the usual form of interactions with an external electromagnetic field $A^{\mu}(x)$ is reproduced by making the $G_{5}$- and gauge invariant replacement:
\begin{equation}
\partial_{\mu} \to D_{\mu} \equiv \partial_{\mu}-(ie/c\hbar)A_{\mu}(x),
\end{equation} 
where $A^{\mu}(x)=(\vec{A},0,\frac{c}{u}A)$ with $\vec{A}$ and $A$ being vector and scalar potentials, respectively, and $\partial_{5}A^{\mu}(x)=0$ is assumed. In this way we find, e.g., the gyromagnetic ratio $g=2$ for the spin 1/2 case [3].

Similarly, the interaction with a given (gravitational) Newton potential $\Phi(\vec{x})$ is introduced into (2) or (4) by making a series of rewriting and replacements : $\eta^{\mu\nu}=h^{\mu}_{a}h^{\nu}_{b}\eta^{ab}, \gamma^{\mu}=h^{\mu}_{a}\gamma^{a}$;
\begin{equation}
h^{5}_{4} \to h^{5}_{4}-\frac{1}{u^{2}}\Phi(\vec{x});
\end{equation}
and then $h^{\mu}_{a}=\delta^{\mu}_{a}$. The potential term thus obtained is $m \Phi(\vec{x})$.

\noindent {\bf b}) In an inertial system S$_{0}$ we introduce by (16) the interaction with $\Phi(\vec{x})=\vec{g}\vec{x}$ ($\vec{g}$: gravitational acceleration), and move afterwards to the system S by (6) with $R(t)={\rm I}$, and $\vec{A}(t)=\frac{1}{2}\vec{a}t^{2}$ ($\vec{a}$: constant vector). The resulting equation for $\psi^{\prime}(\vec{x}^{\prime},t^{\prime})$ or $\psi^{\prime}_{1}(\vec{x}^{\prime},t^{\prime})$ then obtains the potential term $m(\vec{g}-\vec{a})\vec{x}^{\prime}$. Thus, in S with $\vec{a}=\vec{g}$ the gravitational effect completely disappears. The result is basically the same for the case of general $\Phi$'s. This implies that NRQM is compatible with Einstein's equivalence principle.

\noindent {\bf c}) Results from our formalism agree with the non-relativistic, Pauli-, Newton-,$\cdots$ approximations to the corresponding cases of RQM. In fact, our calculations run almost parallel with those of RQM; approximate equations in the latter hold true exactly in the former.

\noindent {\bf d}) Our work shows that the method proposed by Marmo et al. [4] can be extended to those cases in which Lagrangians do not remain invariant after transformations.

\noindent {\bf e}) In our formalism some of the problems which have so far been discussed by invoking classical mechanics become discussible within the framework of NRQM.

\noindent {\bf f}) NRQM is usually regarded as a theory subordinate to RQM, for the former is a special case of, and hence obtainable by approximation from, the latter. In view of the present results, however, we should say that NRQM is an independent theory, being comparable to RQM.

\end{document}